\documentclass[a4paper,11pt]{article}

\newcommand{\rh}[0]{\mathcal{R}_{H}}
\newcommand{\rs}[0]{\mathcal{R}_{BAO}}
\newcommand{\Dr}[1]{\mathcal{D}_{2}^{#1}(r)}
\newcommand{\Dd}[0]{\mathcal{D}_{2}}

\newcommand{\hMpc}[0]{h^{-1}\mathrm{Mpc}}

\newcommand{\hGpct}[0]{h^{-3}\mathrm{Gpc}^3}

\newcommand{\LCDM}[0]{\Lambda CDM}
\newcommand{\LCDMn}[0]{$\Lambda$CDM}
\newcommand{\wknot}[0]{\mathrm{w}}

\newcommand{\refEq}[1]{Eq.~\ref{#1}}
\newcommand{\refF}[1]{Fig.~\ref{#1}}

\newcommand{\refS}[1]{section~\ref{#1}}
\newcommand{\refT}[1]{table~\ref{#1}}
\newcommand{\refApp}[1]{appendix~\ref{#1}}

\newcommand{\citetpi}[1]{(\citet{#1})}

\pdfoutput=1 
             
\usepackage{jcappub} 

\usepackage[T1]{fontenc} 
\usepackage{rotating} 
\usepackage[toc,page]{appendix} 

\title{The scale of cosmic homogeneity as a standard ruler  \\ \today}
\author[a,c,1]{Pierros Ntelis,}
\author[c]{Anne Ealet,}
\author[c]{Stephanie Escoffier,}
\author[a]{Jean-Christophe Hamilton,}
\author[c]{Adam James Hawken,}
\author[b]{Jean-Marc Le Goff,}
\author[b]{James Rich,}
\author[c]{\& Andre Tilquin}

\affiliation[a]{APC, Universit\'{e} Paris Diderot-Paris 7, CNRS/IN2P3, CEA, Observatoire de Paris,\\ 10, rue A. Domon \& L. Duquet,  Paris, France}
\affiliation[b]{IRFU/CEA, Universit\'{e} Paris-Saclay, D36, 91190 Saclay, Gif-sur-Yvette, France}
\affiliation[c]{CPPM, Aix Marseille University, 163, avenue de Luminy, Marseille, France}

\emailAdd{pntelis@cppm.in2p3.fr}
\emailAdd{ealet@cppm.in2p3.fr}
\emailAdd{escoffier@cppm.in2p3.fr}
\emailAdd{hamilton@apc.in2p3.fr}
\emailAdd{hawken@cppm.in2p3.fr}
\emailAdd{jean-marc.le-goff@cea.fr}
\emailAdd{james.rich@cea.fr}
\emailAdd{atilquin@cppm.in2p3.fr}

\abstract{
In this paper, we study the characteristic scale of transition to cosmic homogeneity of the universe, $\rh$, as a standard ruler, to constrain cosmological parameters on mock galaxy catalogues. We use mock galaxy catalogues that simulate the CMASS galaxy sample of the BOSS survey in the redshift range $0.43 \leq z \leq 0.7$. In each redshift bin we obtain the homogeneity scale, defined as the scale at which the universe becomes homogeneous to $1\%$, i.e. $D_2(\rh) = 2.97$. 
With a simple Fisher analysis, we find that the performance of measuring the cosmological parameters with either the position of the BAO peak or the homogeneity scale is comparable. We show that $\rh$ has a dependence on the galaxy bias. If the accuracy and precision of this bias is achieved to $1\%$, as expected for future surveys, then $\rh$ is a competitive standard ruler.  

}


\begin{document}
\maketitle
\date
\flushbottom

\section{Introduction}\label{Introduction}

The standard model of cosmology, known as flat $\Lambda CDM$, describes a Universe mainly composed of Cold Dark Matter (CDM) and a cosmological constant $\Lambda$.  The two main assumptions of this model are the validity of General Relativity as an accurate description of gravity and the {\em Cosmological Principle}~\citetpi{CP} that states that the Universe is isotropic and homogeneous on large enough scales. This model shows excellent agreement with current data, be it from type Ia supernovae~\citetpi{SnIaPerlmutter,SnIaRiess,betoule2014,2018arXiv181102374D}, temperature and polarisation anisotropies in the Cosmic Microwave Background~\citetpi{aghanim2018planck} or Large Scale Structure Clustering~\citetpi{LssPercival,LssParkinsonWg,LssHeymans2013Cfhtlens,BAOcosmo,2017MNRAS4702617A}.

In \citet{ntelis2017exploring} (henceforth N17), we studied the characteristic scale of transition to cosmic homogeneity of the universe, the homogeneity scale, assuming the standard model of cosmology. Historically, the concept of homogeneity in the large scale structure of the Universe can be traced back to \citet{1760pnpm.book.....N}. In the modern era, \citet{Martinez2} were the first to measure the homogeneity scale in the sky, suggesting a homogeneity scale larger than $100\ h^{-1}$Mpc.  Since then, several methods have been developed to study the homogeneity scale \cite{Hogg,Yadav,Sarkar,Guzzo,Martinez2,Scaramella,Amendola,PanColes,kurokawa2001scaling,WiggleZ,2018arXiv180911125G,avila2018scale}. In this work, we follow the method first proposed by \citet{WiggleZ} and further developed by N17. However, what we really measure in these studies is the combination between the volume distance in the fiducial cosmology and a characteristic scale, similar to BAO studies  \citetpi{2015A&A...584A..69R}. Therefore, we will present the results from N17 divided by the volume distance in the fiducial cosmology.

Objects with a characteristic luminosity, such as type Ia supernovae, can be used as standard candles to probe cosmology \citetpi{SnIaPerlmutter,SnIaRiess,betoule2014}. Likewise, characteristic scales in the statistics of the clustering of galaxies, such as the position of the BAO peak \citetpi{eisenstein2005detection} can be used as standard rulers. Standard rulers are important in cosmology since they allow us to measure cosmological distances as a function of redshift. The relationship between distance and redshift is dictated by the rate of cosmic expansion and curvature. Thus, by studying standard rulers, we can improve upon our understanding of cosmology.

In this paper, we perform a proof of concept study. Using mock galaxy catalogues, we demonstrate that the homogeneity scale can be used as a standard ruler to constrain cosmology. We define the homogeneity scale as the scale at which the galaxy distribution is homogeneous to 1\%, according to its fractal properties. A full definition of what we mean by "homogeneous" is given in \refS{sec:Methodology}.

This document is structured as follows: In \refS{sec:Dataset}, we describe the mock galaxy catalogues. In \refS{sec:Methodology}, we describe the use of the homogeneity scale as a standard ruler and we compare it with the BAO standard ruler, i.e. the position of the BAO peak. In \refS{sec:Cosmology}, we explain how we can extract cosmological information from this new standard ruler. Finally, in \refS{sec:conclusion}, we discuss our conclusions.

\section{Mock galaxy catalogues}\label{sec:Dataset}

In this study, we use 1000 Quick Particle Mesh (QPM) mock galaxy catalogues \citetpi{QPM} designed to simulate the CMASS galaxy sample of the Baryon Oscillation Spectroscopy Survey (BOSS) \cite{dawson2012baryon,BOSS-DR12,eisenstein2011sdss}. Each mock catalogue has a sky coverage of $10,400$ deg$^2$. Objects were selected following the CMASS colour cuts described in \citet{TargetSelection}. 
We selected objects in the redshift range of $0.43<z<0.7$.  For this study, we used only the North Galactic Cap (NGC). The NGC has a larger area compared with the South Galactic Cap which allows us to measure our observable more precisely, as we have shown in N17. NGC has an area $\sim7000$ deg$^2$.
The flat $\LCDM$ cosmology used to obtain these catalogues is given by:
\begin{equation}\label{fid-cosmo}
	\textbf{p}_{\mathrm{F}}=(h,\omega_{b},\omega_{cdm},n_{s},\ln\left[10^{10}A_{s}  \right] ,\Omega_k) = (0.7, 0.0225,0.11172,0.95,3.077,0.0) \; .
\end{equation}


where $h=H_0/[100$ km s$^{-1}$ Mpc$^{-1}]$ is the dimensionless Hubble constant with $H_0$ the  Hubble constant, $\omega_{b} = \Omega_{b}h^{2}$, is the reduced baryon density ratio, $\omega_{\rm cdm} = \Omega_{\rm cdm}h^{2}$ is the reduced cold dark matter density ratio, $n_{s}$  the spectral index, $A_{s}$ the amplitude of the primordial scalar power spectrum and $\Omega_k$ is the curvature density ratio. In this framework, the Dark Energy density ratio is defined via $\Omega_{\Lambda} = 1 - \Omega_m - \Omega_k$, where $\Omega_m$ is the total matter density ratio.

This defines the fiducial cosmology\footnote{This fiducial cosmology is in agreement with \citet{aghanim2018planck}}, that we are using, to convert the $z,R.A.,Dec$\footnote{ The R.A. and Dec are the abbreviations of Right Ascension and Declination measured in degrees.} information in the catalogue, into comoving coordinates, using the comoving distance relation, see \refApp{sec:Cosmography}. This gives an effective volume of $V_{\mathrm{eff}} \simeq 3\ \hGpct$, for the entire survey. However, this fiducial cosmology biases the $\rh^{fid}$ towards itself. Therefore, we need to correct for this effect as we explain in \refS{sec:MCMC}. Note that a simple extension of this cosmology is a time varying Dark Energy density ratio, parametrized by $\wknot$. This extension allows us to investigate models of modified gravity \cite{PIERROS_THESIS,montanari2015measuring} and we explore this extension in \refS{ch:FisherMat}. 


\section{Methodology}\label{sec:Methodology}

In this section, we describe how we can use the homogeneity scale, $\rh$, as a standard ruler. Following N17 (and \citet{WiggleZ}), we use the fractal dimension, $\mathcal{D}_{2}(r)$, as a metric of homogeneity and estimate this in the mock galaxy catalogues. Our observable is the fractal dimension. The fractal dimension is related to the counts-in-spheres $N(<r)$ according to $N(<r) \propto r^{D_2}$. For a completely homogeneous distribution $\mathcal{D}_2=3$. While for a fractal distribution, it deviates from this value. Our observable is related to the two-point correlation function according to:
\begin{equation}\label{eq:d2observable}
	\mathcal{D}_2(r) = 3 + \frac{d\ln }{d\ln r} \left[  1+ \frac{3}{r^{3}} \int_{0}^{r}{\xi}(s)s^{2}ds  \right] \; ,
\end{equation}
where $\xi(s)$ is the usual two-point correlation function that it is used in large scale structure studies\footnote{This observable is calculated using a publicly available code \url{https://github.com/lontelis/cosmopit} }. Now we are able to construct a standard ruler according to a characteristic scale of homogeneity, as:
\begin{equation}\label{eq:RH-definition}
 \mathcal{D}_{2}(\rh) = 2.97 \; .
\end{equation}
This defines the scale at which the Universe becomes homogeneous to within 1\%. 

We use the perturbed einstein-boltzman equations (implemented in  CLASS\citetpi{CLASS}, which includes implicit assumption on the primordial power spectrum and the matter transfer function) to compute the theoretical matter power spectrum, $P(k)$, for our fiducial cosmology. Applying a fourier transform, we get $\xi(r)$. Then using \refEq{eq:d2observable} and \refEq{eq:RH-definition}, we compute the fractal dimension and the homogeneity scale, respectively. 



We extracted the homogeneity scale, $\rh$ and the position of the BAO peak, $\rs$, from the same mock galaxy catalogues for comparison purposes.
We studied the homogeneity scale using the galaxy distribution rather than the total matter distribution. Therefore, we need to take into account the bias in the final analysis as we explain in \refS{sec:MCMC}. 

\begin{table}[h!]
		\begin{center} 
		\begin{tabular}{c|c|c} 
		\hline
		 & $\mathcal{R}_{H}^{fid} / d_V^{fid}  $& $d_V^{fid} [h^{-1}\mathrm{Mpc}]$\\
		\hline 
		$0.430-0.484$ 	& $0.108 \pm 0.007$  &  1165 \\
		$0.484-0.538$ 	& $0.094 \pm 0.005$  &   1275 \\
		$0.538-0.592$ 	& $0.089 \pm 0.005$  &   1381 \\
		$0.592-0.646$  & $0.083 \pm 0.005$ &   1482 \\
		$0.646-0.700$  & $0.082 \pm 0.005$  &   1577 \\

	 	\hline
		\end{tabular}
		\end{center}
\caption{\label{tab:RH-z} Mean and standard deviation of the normalised homogeneity scale, $\rh^{fid}/d_V^{fid}$, as a function of redshift, $z$, for the galaxy distributions in the north galactic cap (NGC), for the 1000 QPM mock catalogues, as explained in \refS{ch:Rh_determination}. The last column is the fiducial volume distance, $d_V^{fid}$ \refApp{sec:Cosmography}. }
\end{table}

\subsection{Estimation of $\rh$ and $\rs$}\label{ch:Rh_determination}

We first used the Landy $\&$ Szalay estimator \cite{LSestimator} to measure the two-point correlation function in the mock galaxy catalogues. Then in the range of $r=[40,180]\hMpc$, we fitted the two point correlation function around the position of the BAO peak, following \citet{anderson2012clustering}, see \refApp{sec:BAOpeakmethodology}.

	Using \refEq{eq:d2observable}, we integrated the two point correlation function and we obtained the fractal dimension as a function of scale. To estimate the homogeneity scale, we fit a spline function to $\Dd{}$ (\refEq{eq:d2observable}) over the range $r=[90,200] \hMpc$.  Using the definition in \refEq{eq:RH-definition}, we then obtained the homogeneity scale,  $\rh^{fid}$ in the fiducial cosmology \footnote{We also used a polynomial fit to estimate this scale over the ranges, $r=[10,1300] \hMpc$, and we found no significant disagreement.}. 
We took the values of $\rh^{fid}$, in the redshift range $0.430 < z < 0.700$ from N17. However, these values are measured using the fiducial cosmology. Therefore we note the values with the normalisation according to the volume distance, $d_V$ (see \refApp{sec:Cosmography}). We show the results in \refT{tab:RH-z}.

		\begin{figure}[ht!]
	    \centering 
	    \includegraphics[width=140mm]{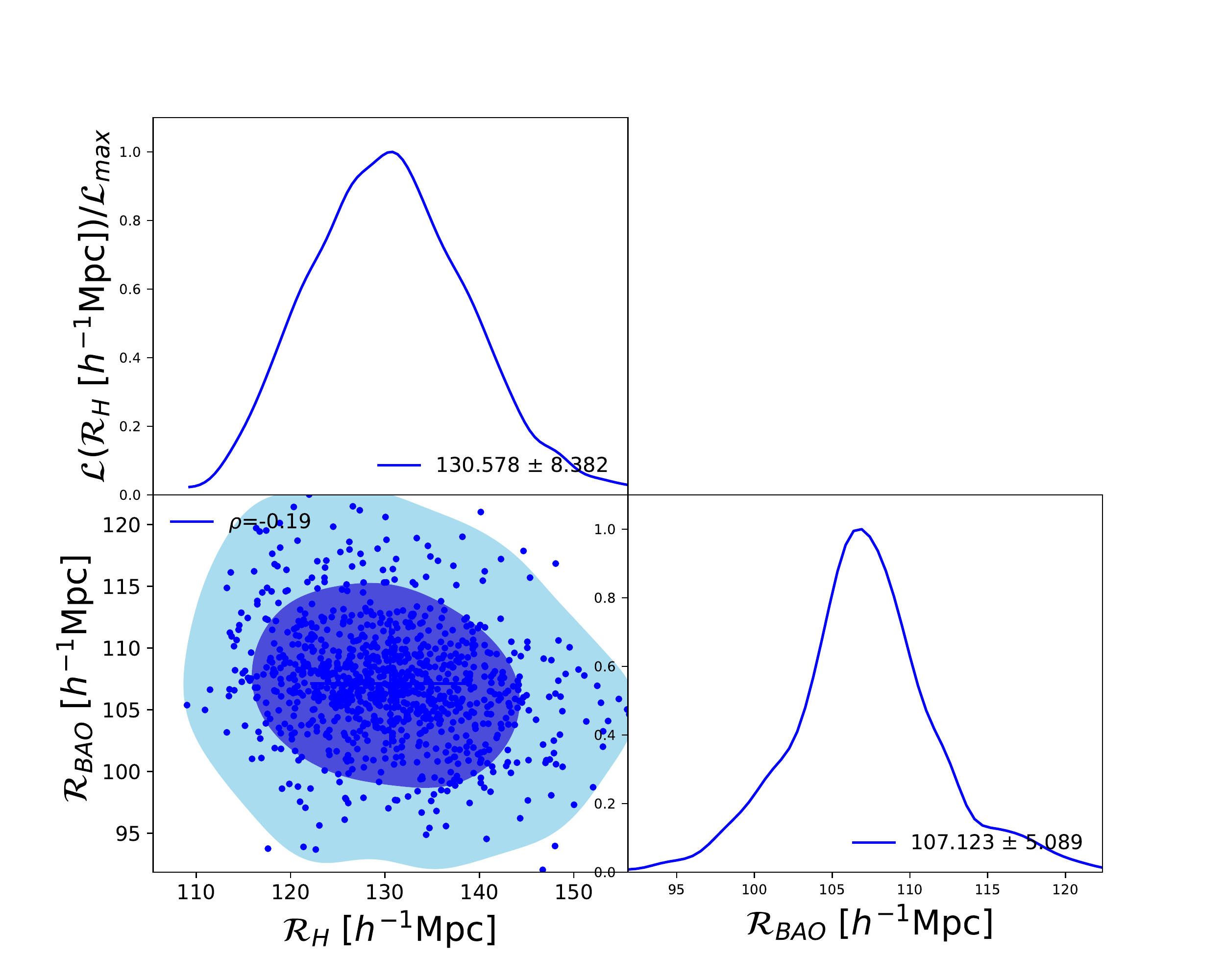}
	    \caption{\label{fig:RH_BAO_contour} Contour plot with $1\sigma$ ($2\sigma$) in dark blue (light blue) for the Homogeneity scale versus the position of the BAO peak for the $0.538 \leq z < 0.592$, using 1000 mock catalogues. There is negligible correlation between the two scales, $|\rho|< 0.20$. The number on the top right up corner of the panels correspond to mean and the standard deviation value of the observables. }
	   \end{figure}

 In \citet{PIERROS_THESIS}, we have shown that the homogeneity scale has no dependence on the observational systematic effects of our survey. 

\subsection{Correlation of $\rh-\rs$ estimates}\label{ch:CrossCorrelationRhRs}
	
	In this section, we study the correlation coefficient between the estimate of the Homogeneity scale and the estimate of the position of the BAO peak,~$\rho=C_{\rh \rs}/\sqrt{\sigma_{\rh}\sigma_{\rs}} $, where $C$ denotes the covariance of the two scales and $\sigma$ denotes the standard deviation of each scale. 

	To study the correlation, we estimated the two standard rulers in $1000$ QPM mock catalogues. We measured the correlation between the Homogeneity scale and the position of the BAO peak both determined via the methods as described above. We found only a small correlation between the two scales, for example, in the redshift bin $0.538 \leq z < 0.592$, $\rho\simeq -0.19$, as shown in \refF{fig:RH_BAO_contour}. We find similar results in the rest of the redshift bins. Explicitly, we find that $|\rho|<0.3$, for all redshift bins.	

The Homogeneity scale and the position of the BAO peak both have units of $\hMpc$ and are of a similar magnitude but the fact that the correlation coefficient between the two is $|\rho|<0.3$ means that they are fairly independent of one another. Therefore, we can investigate the use of the homogeneity scale as an independent standard ruler.
	   	   
\section{Cosmology with $\rh$}\label{sec:Cosmology}
We implement two techniques to assess the performance of $\rh$ and $\rs$ as standard rulers to constrain cosmological parameters. We proceed with the two following steps. First, we perform a Fisher analysis to investigate the sensitivity of $\rh$ and $\rs$ to cosmological parameters  \citetpi{2006DEFT}. However, this is a Gaussian approximation of estimating the errors. Therefore, in the second step, we perform a Monte Carlo Markov Chain (MCMC) analysis to investigate the performance of the two probes as standard rulers in mock galaxy catalogues. 

\subsection{Fisher analysis}\label{ch:FisherMat}
The Fisher matrix is defined by the derivatives of an observable at different redshifts, $O(z)$, as a function of the parameters, $p_{i}$, that this observable depends upon \citetpi{2006DEFT}. This is defined as:
\begin{equation}
	F_{ij} = \sum_{z} \frac{1}{\sigma_O^2(z)}\frac{\partial O(z)}{\partial p_{i} } \frac{\partial O(z)}{\partial p_{j} }
\end{equation}
where $\sigma_O(z)$ is the error on the observable at each redshift.
For simplicity, we have assumed the same precision of each observables in all redshift bins. A simple extension to the \LCDMn-model, as discussed in \refS{Introduction}, is one with a varying Dark Energy density, i.e. the $\mathrm{w}\mathrm{CDM}$-model. Therefore, we look at the following parameters:
\begin{equation}
	\textbf{p}_{\mathrm{w}\mathrm{CDM}} = \left\{ h,n_s,\omega_{b},\omega_{cdm},n_s,\ln 10^{10}A_s,\Omega_{\Lambda},\mathrm{w}\right\}
\end{equation}
and compare the amount of information on them gained by using the standard rulers, $\rh$ and $\rs$. From observations we have a linear dependence on the cosmic linear bias, $b$. This parameter is degenerate with the $A_s$ parameter via $b^2A_s$, therefore we do not consider it here. However, in \refS{sec:MCMC}, we take bias into account.

We find that the Homogeneity scale improves cosmological constraints, when used in combination with the CMB \citetpi{Params-Planck}. These constraints are comparable to those of the position of the BAO peak combined with the CMB. The constraints on $A_s$, $\Omega_{\Lambda}$, $\wknot$, $h$ are significantly improved, while the contribution to $\omega_{cdm}, \omega_{b}$ and $n_s$ is negligible for both probes\footnote{See Fig 1 of \citet{Ntelis:2018tlj} proceedings. }. As we present in \refT{tab:FisherPrecision}, for $\Omega_{\Lambda}$, $\rs$ provides better constraints than $\rh$. When we look at $\mathrm{w}$, $\rh$ provides comparable constraints to $\rs$.

\begin{table}[ht!]
		\begin{center} 
		\begin{tabular}{c|ccc} 
	 	      $68\%$               &  CMB & $+\rh$ & $+\rs$ \\
		\hline 
		$\sigma_{\Omega_{\Lambda}}$ 	& 0.06  & 0.04 & 0.03  \\
		$\sigma_{\mathrm{w}}$ 	& 0.30  & 0.20 & 0.20  \\
	 	\hline
		\end{tabular}
		\end{center}
\caption{\label{tab:FisherPrecision} Precision of $1\sigma$ (68\% C.L.) of the $\Omega_{\Lambda}-\wknot$ plane from the Fisher analysis, for the different probes, the CMB, and the combination of CMB with the homogeneity scale, $\rh$, or the position of the BAO peak, $\rs$. [see text for details]}
\end{table}

The homogeneity scale is sensitive to $\Omega_{\Lambda}$ and $h$ since
an accelerating expansion damps the growth of structures, rapidly decreasing the homogeneity scale. However, the expansion rate and the acceleration are two correlated phenomena. Therefore, we can constrain only one of them. In our case we choose $\Omega_{\Lambda}$. These results show that given the improvement on $\Omega_{\Lambda}$, the homogeneity scale can be used to explore external models such as non-flat universes or universes with variable Dark Energy equation of state. The Fisher analysis also suggests that the constraints from the homogeneity scale and the position of the BAO peak are not orthogonal and so the combination of $\rh$ and $\rs$ does not provide significantly improved constraints relative to $\rs$ alone. 


\subsection{MCMC analysis}\label{sec:MCMC}
From the Fisher analysis we learnt that the homogeneity scale is sensitive to $\Omega_{\Lambda}$ providing negligible constraints on the other parameters of the model. Therefore, in this section, we implement an MCMC analysis to determine the constraints provided by the homogeneity scale in the cosmological parameter space $p_C=(\Omega_{\Lambda},\Omega_{m})$. This is the open-$\Lambda\mathrm{CDM}$ model. 

The $\chi^2$ that we explore with an MCMC algorithm\footnote{We use the publicly available code, pymc \url{https://pymc-devs.github.io/pymc/}. } is given by:
\begin{equation}\label{eq:chi2_final}
		\chi^2 (A,B,p_C) = \sum_{z=1}^{5} \left( \frac{\rh^{G}(z;p_{F}) - \rh^{M,th}(z;p_F) \times b_{\rh}(z;A,B) \times \alpha(z;p_{C},p_F) }{ \sigma_{\rh^{G}}(z) }  \right)^{2}
\end{equation}
where $p_F$ are the cosmological parameters fixed to their fiducial values; $\rh^{G}(z;p_{F})$ is the homogeneity scale of the galaxy distribution as a function of redshift as measured in the fiducial cosmology; $\sigma_{\rh^{G}}(z)$ is the error on $\rh^{G}(z;p_{F})$; $\rh^{M,th}(z;p_M)$ is the prediction of the homogeneity scale of the matter distribution as a function of redshift; $\alpha(z;p_{C},p_F)=d_V(z;p_C)/d_V(z;p_F)$ is the ratio of the volume distance in a given cosmology to the fiducial value. The homogeneity scale of the galaxy distribution is biased with respect to the homogeneity scale of the total matter distribution. To account for this bias, we use a linear bias model, originally designed for the two point correlation function \citetpi{basilakos2008halo}, 
adapted to fit the homogeneity scale: 
\begin{equation}
	b_{\rh}(z;A,B) = A\left( \frac{1+z}{1+z_{eff}} \right) ^B \; ,
\end{equation}  
where $z_{eff}$ is the value of the intermediate redshift bin (where our estimate of A is the most accurate). The factor $(1+z_{eff})$ is included to reduce the degeneracy between A and B parameters\footnote{Notice that this is a new way of parametrising the linear galaxy bias. }.

We estimated the homogeneity scale on the mock galaxy catalogues, as described in \refS{sec:Dataset}. We measured the mean and standard deviation of these values. We then performed an MCMC using the mean of the mocks as our data and the standard deviation as our error. First we fixed the cosmological parameters to their fiducial values to find the best fitting values of the bias parameters in our cosmology, i.e. that of the mock galaxy catalogue. We find $(A,B)=(1.975\pm0.052,0.999\pm0.597)$. The A parameter is close to the normal linear galaxy bias.


	   \begin{figure}[h!]
	   \hspace{-1.cm} 
	    \includegraphics[width=160mm]{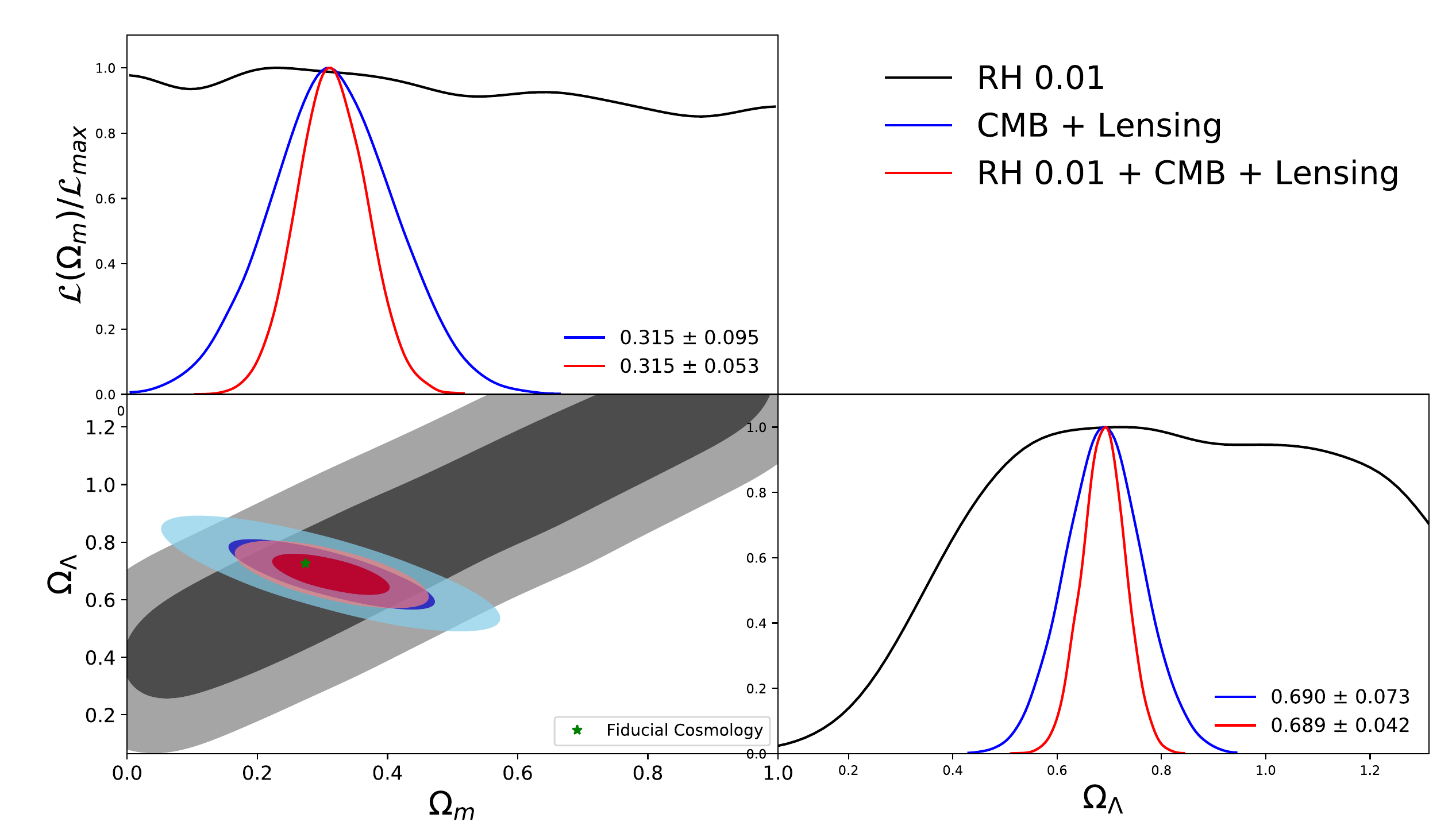}
	    \caption{\label{fig:QPM_Prior_AB_RH_RHCMB_RHCMBLensing001_unnormalized_om_ol} Contours of 68\% (dark) and 95\% (light) of the $(\Omega_m,\Omega_{\Lambda})$ plane using $\rh$ (black), CMB+Lensing (blue), $\rh$+CMB+Lensing. The green star denotes the values of the fiducial cosmology of the mock galaxy catalogues. The diagonal panels show the normalised likelihood and the mean and the standard deviation of each parameter colour-coded for each probe. The precision on bias parameters is 1\%. [see text for details].}
	   \end{figure}%

We then freed our cosmological parameters and ran the MCMC with the addition of Gaussian priors on A and B centred on our previous estimates, and priors from the CMB+Lensing \citetpi{Params-Planck}\footnote{We extract graphically the $(\Omega_m,\Omega_{\Lambda}$)-information  from Fig. 26 page 38 of \citetpi{Params-Planck}}. We assessed the ability of priors of different widths (on A and B) to recover our fiducial cosmology in order to infer the precision to which our bias parameters must be known. We found that in order to improve upon the constraints from the CMB+Lensing alone the required precision is less than $2\%$. We then performed an MCMC analysis by applying a $1\%$ prior to our bias parameters. 
The precision and accuracy considered for this bias parametrisation is not obtainable from current observations. However, it will be obtainable from future surveys such as Euclid \citetpi{amendola2017constraints}. 
In \refF{fig:QPM_Prior_AB_RH_RHCMB_RHCMBLensing001_unnormalized_om_ol}, we show the marginalised contours of the $(\Omega_m,\Omega_{\Lambda})$ plane for $\rh$ alone (with 1\% precision on the bias parameters A and B) (black), CMB+Lensing (blue), the combination of $\rh$+CMB+Lensing (red). The green star denotes the values of the fiducial cosmology used to generate the mock galaxy catalogues. The addition of information from $\rh$ improves constraints relative to the CMB+Lensing alone by a factor of 0.56 
for $\Omega_{m}$
and 
0.56
for $\Omega_{\Lambda}$. We find, the mean of the marginalised likelihood of $\Omega_{m}^{\mathrm{\rh + \mathrm{CMB} +\mathrm{Lensing} }} =0.318  \pm 0.054$
and $\Omega_{\Lambda}^{\mathrm{\rh + \mathrm{CMB}+\mathrm{Lensing} }} =0.688 \pm 0.042$~. When we compare 
\refF{fig:QPM_Prior_AB_RH_RHCMB_RHCMBLensing001_unnormalized_om_ol} with, for example, \citet{bautista2017measurement}, we see that the orientation of the constraints provided by $\rh$ is comparable to the orientation of those obtained using BAO measurements, using galaxies. However, considering the fisher analysis, described in \refS{ch:FisherMat}, constraints obtained from the two standard rulers are not orthogonal to one another, limiting their use in combination.

This demonstrates that $\rh$ can be used as a standard ruler to recover the input cosmology. Thus, $\rh$ can be used as a standard ruler to constrain cosmological parameters. In particular, it can be used to improve the measurement of the $(\Omega_m,\Omega_{\Lambda})$ plane\footnote{Our analysis is available under GNU licence \url{https://github.com/lontelis/CoHo}. }.


\section{Conclusion and Discussion}\label{sec:conclusion}

We have demonstrated that the characteristic scale of transition to cosmic homogeneity, $\rh$ can potentially be used as a standard ruler to probe cosmology with large scale structure surveys. 

We have compared the precision of cosmological parameters obtained using $\rh$ with those obtained with the position of the BAO peak, $\mathcal{R}_{BAO}$, using mock galaxy catalogues. We have found that there is only a small correlation between the two probes, $|\rho|< 0.30$, making the homogeneity scale a complementary cosmological probe. 

In order to quantify the additional information contained in the homogeneity scale, we have performed a simple Fisher analysis with a set of cosmological parameters of the flat wCDM model. We have found that the homogeneity scale gives comparable information to that of the position of the BAO peak for this set of cosmological parameters. In particular, the homogeneity scale is sensitive to the dimensionless Hubble constant, $h$, the amplitude of the primordial fluctuations, $A_s$, the Dark energy density ratio, $\Omega_{\Lambda}$, and the equation of state, $\wknot$. This shows the implicit dependence of the homogeneity scale on the shape of the transfer function. However, constraints obtained from the two standard rulers are not orthogonal to one another, limiting their use in combination.

Using an MCMC algorithm, and applying a $1\%$ prior to our bias parameters, we explored the open $\LCDM\ $ model on the mean of the mock galaxy catalogues. We found 
~$\Omega_{m}^{\mathrm{\rh+\mathrm{CMB}+\mathrm{Lensing} }} =0.318  \pm 0.054$
and
~$\Omega_{\Lambda}^{\mathrm{\rh+\mathrm{CMB}+\mathrm{Lensing} }} =0.688 \pm 0.042$,  consistent with the input flat \LCDMn-model cosmology of the mock galaxy catalogues. The inclusion of $\rh$ improves CMB$+$Lensing constraints by a factor of 0.56 for the total matter density ratio and by a factor of 0.56 for Dark Energy density ratio constraints. These results show the sensitivity of our probe to cosmic bias.

In summary, we have shown the dependence of the homogeneity scale to the matter power spectrum showing the parameter dependence. In a future study, we are going to investigate the observational selection effects dependence of our probe\cite{ntelisFuture}. Therefore, we have revealed the complementarity of the homogeneity scale with respect to other cosmological probes. 

Finally, we stress that this analysis can be performed and improved upon in the light of more observational data from current and future experiments such as SDSS-IV (eBOSS) \citetpi{dawson2016sdss}, Euclid \citetpi{2016arXiv160600180A}, LSST \citetpi{2009arXiv0912.0201L} and DESI \citetpi{aghamousa2016desi}. Furthermore, analogous methods could be applied to data from SKA \citetpi{dewdney2009square}. We relegate this analysis to future work.

\textbf{Note added:} A recent paper \citetpi{2018arXiv180911125G} appeared simultaneously with this work. They measured the homogeneity scale in the eBOSS DR14 QSO sample with a similar methodology to the one presented here, but at a higher redshift, $0.80<z<2.24$. They acquired a similar precision to our measurement on mocks, but on real data. Therefore, we can apply our analysis, the homogeneity scale as a standard ruler, to their measurement. On the other hand, another recent paper \citetpi{2018arXiv181003539G}, appeared simultaneously with this work. They measured several quantities related to the fractal dimension in a galaxy catalogue of SDSS-DR7 at lower redshifts $z<0.5$. They find values that do not agree with the homogeneity scale found and used in this work. Thus, we do not expect that our method will give reliable results to this chosen galaxy catalogue. However, several updates on the construction of the galaxy catalogue in their redshift bin have been made since then \citetpi{TargetSelection}.


\vspace{0.5cm}
\hspace{-1cm} \textbf{AKNOWLEDGEMENTS}\\
	We would like to thank the anonymous referee, for his constructive comments that improved our manuscript.
	
	PN would like to thank Jean-Paul Kneib, Marc Lachieze-Ray, Stavros Katsanevas, Stéphane Plaszczynski, Nicolas G. Busca, Cyrille Doux, Mikhail Stolpovskiy, Hector Gil Marin and Charling Tao for useful suggestions and discussions. 
		
	PN was funded by Ecole Doctoral 560, university Paris Diderot for the early stages of this analysis and currently he is funded by  'Centre National d'\'etude spatiale' (CNES), for the Euclid project.
	
	AJH acknowledges the financial support of the
OCEVU LABEX (Grant No. ANR-11-LABX-0060) and
the A*MIDEX project (Grant No. ANR-11-IDEX- 0001-
02) funded by the Investissements d'Avenir French government program managed by the ANR. This work also acknowledges
support from the ANR eBOSS project (ANR-16-CE31-0021) of the French National Research
Agency.
	
	This research used also resources of the IN2P3/CNRS and the Dark Energy computing Center funded by the OCEVU Labex (ANR-11-LABX-0060).
	
	The French Participation Group of SDSS-III was supported by the Agence Nationale de la Recherche under contracts ANR-08-BLAN-0222 and ANR-12-BS05-0015-01.
	
	Funding for SDSS-III has been provided by the Alfred P. Sloan Foundation, the Participating Institutions, the National Science Foundation, and the U.S. Department of Energy Office of Science.
The SDSS-III web site is http://www.sdss3.org/.
	
	SDSS-III is managed by the Astrophysical Research Consortium for the Participating Institutions of the SDSS-III Collaboration including the University of Arizona, the Brazilian Participation Group, Brookhaven National Laboratory, Carnegie Mellon University, University of Florida, the French Participation Group, the German Participation Group, Harvard University, the Instituto de Astrofisica de Canarias, the Michigan State/Notre Dame/JINA Participation Group, Johns Hopkins University, Lawrence Berkeley National Laboratory, Max Planck Institute for Astrophysics, Max Planck Institute for Extraterrestrial Physics, New Mexico State University, New York University, Ohio State University, Pennsylvania State University, University of Portsmouth, Princeton University, the Spanish Participation Group, University of Tokyo, University of Utah, Vanderbilt University, University of Virginia, University of Washington, and Yale University.

\begin{appendices}

\section{Cosmography}\label{sec:Cosmography}
From $z,R.A.,Dec$ we need to infer distances. Therefore we reconstruct the $z$ information according to standard cosmology and then we make our measurements in comoving space. In standard cosmology, we define the following distances.

The \textit{comoving distance}:
\begin{equation}\label{eq:FRW}
		d_{\rm C} (z) =c \int_{0}^{z} \frac{dz'}{H(z')} \;,
	\end{equation}
where $c$ is the speed of light and
$$\label{eq:Hz}
	H(z) = H_{0} \sqrt{ (\Omega_{\rm cdm} +\Omega_{\rm b} )(1+z)^{3} + \Omega_{\Lambda}  +\Omega_k(1+z)^2 } \; ,
$$ 
is the usual Hubble expansion rate.

We define the \textit{volume distance}:
\begin{equation}\label{eq:d_V}
	d_V(z) = \left[ cz H^{-1}(z) d^2_M(z) \right]^{1/3} \; ,
\end{equation}
where $d_M$ is the \textit{motion distance}:
\begin{equation}
		d_M(z) = 
		\left\{ 
			\begin{matrix}
			\frac{d_H}{\sqrt{\Omega_k}} \sinh \left(  \sqrt{\Omega_k} \frac{d_C(z)}{d_H} \right) \  , \ \Omega_k > 0  \\
			d_C(z) \ , \ \Omega_k = 0 \\
			\frac{d_H}{\sqrt{\Omega_k}} \sin \left(  \sqrt{\Omega_k} \frac{d_C(z)}{d_H} \right) \ , \ \Omega_k < 0						
			\end{matrix} 
		\right\} 
		\; , 
\end{equation}
where $d_C$ is given by \refEq{eq:FRW} and $d_H=c/H_0$.

\section{Determination of the position of BAO peak}\label{sec:BAOpeakmethodology}
	 We used two methods to determine the position of the BAO peak. For the first method,
we model the 2-point correlation function in such a way that the result is only driven by the position of the BAO peak~\citetpi{anderson2012clustering}. 
	We then apply a broadband model. This model can be described by the following formula:  
	\begin{equation}
		bb(r) = p_1 + \frac{p_2}{r} + \frac{p_3}{r^{3}}
	\end{equation}
	where $(p_1,p_2,p_3)$ are the broadband parameters. 
	
	We model the measurement of the position of the BAO peak using the usual Gaussian model \citetpi{sanchez2011tracing}, described by:
	\begin{equation}\label{eq:BAO1}
		\xi^{(1)}(r)  = A\ exp\left[ -\frac{1}{2}\left(\frac{r-\rs}{\sigma_{peak}}\right)^2 \right] + bb(r)
	\end{equation}
	where $\rs$ is the position of the BAO peak parameter, A and $\sigma_{peak}$ are the amplitude and the smoothing scale of a Gaussian function, respectively.

	In order to determine the position of the BAO peak, we measure the parameter $\rs$ by marginalising over the rest of the parameters. We marginalise the nuisance parameters, $(A,\sigma_{peak},p_{bb})$.

	For the second method, we follow the same steps but now we use the usual correlation function with the fiducial cosmology as a template, $\xi(r;p_F)$, instead the Gaussian model. We model the correlation function as:
	\begin{equation}\label{eq:BAO2}
		\xi^{(2)}(r)  = b^{2} \xi(\alpha_{iso}*r;p_F) + bb(r;p_{bb})
	\end{equation}
	where we model the \textit{isotropic dilatation} parameter as:
	\begin{equation}
		\alpha_{iso} = r_{s} / r_{s}^{fid} \; .
	\end{equation}
	Note that now $\rs=r_s$. In this case the nuisance parameters are $(b,p_{bb})$. The two methods, explained above, provide similar conclusions.

\end{appendices}



\label{Bibliography}


\bibliographystyle{unsrtnat_arxiv} 

\bibliography{Bibliography} 

\end{document}